\RequirePackage{ifpdf}
\documentclass{PoS}

\usepackage{graphicx}
\usepackage{amsmath}
\usepackage{amssymb}
\usepackage[mathscr]{eucal}
\usepackage{fancyvrb} % nach \begin{document} Eintrag: \VerbatimFootnotes
\usepackage[hyphens]{url} % allows for linebreaks in: \url{http://www.long.name.and.another.name.pl}.
\usepackage[pdftex]{color}     % forgot the role of [pdftex]
\let\normalcolor\relax         % 2010-03-14 - needed for making pdflatex compatible with package color
\def \ar      {\texttt{AMBRE}{}}

\def \la {\texttt{LA}{}}
\def \ga {\texttt{GA}{}}
\def \mb {\texttt{MB}{}}

\title{%
{\flushright{ 
\small \texttt{DESY 14-125}
\\
\small \texttt{DO-TH 14/15}
\\
\small \texttt{LPN 14-089}
\\
\small \texttt{SFB/CPP-14-42}
\\[8mm]
}}
Non-planar Feynman integrals, Mellin-Barnes representations, multiple sums}

\ShortTitle{Non-planar Feynman integrals, Mellin-Barnes representations, multiple sums }

\author{Johannes Bl\"umlein$^a$, Ievgen Dubovyk$^a$, 
\speaker{Janusz Gluza}$^{,b}$, Micha{\l} Ochman$^a$, \newline Clemens G. Raab$^a$,
Tord Riemann$^a$, Carsten Schneider$^c$
\\
\\
\llap{$^a$}  Deutsches Elektronen-Synchrotron, DESY, Platanenallee 6, D-15738 Zeuthen, Germany \\
\llap{$^b$} Institute of Physics, University of Silesia, Uniwersytecka 4, PL-40007 Katowice, Poland \\
\llap{$^c$} Research Institute for Symbolic Computation, J. Kepler University Linz, A-4040 Linz, Austria\\

E-mails: \email{johannes.bluemlein@desy.de}, \email{e.a.dubovyk@gmail.com}, \email{janusz.gluza@us.edu.pl},
\email{michal.ochman@desy.de}, \email{clemens.raab@desy.de}, \email{tord.riemann@desy.de}, \email{cschneid@risc.jku.at}
}

\abstract{The construction of Mellin-Barnes (\mb{}) representations  for non-planar Feynman diagrams and the 
summation of multiple series derived from general \mb{} representations are discussed. A basic version of a new package \ar v.3.0 is 
supplemented.
The ultimate goal of this project is the automatic evaluation of \mb{} representations for multiloop
scalar and tensor Feynman integrals through infinite sums, preferably with analytic solutions.
We shortly describe a strategy of further algebraic summation. 
}

\FullConference{Loops and Legs in Quantum Field Theory - LL 2014, \\
                27 April - 2 May 2014            \\
                Weimar, Germany}

\begin{document}

\section{Introduction}
There are many methods to calculate Feynman integrals, both numerically and analytically. 
Here, we follow an attractive method, relying on changing integrals over Feynman parameters into complex path 
integrals using the Mellin-Barnes (MB) identity
\cite{zbMATH02640947,Usyukina:1975yg,Boos:1990rg,Smirnov:1999gc,Tausk:1999vh,mbtools,Czakon:2005rk,Smirnov:2009up,Gluza:2007rt,%
Gluza:2010rn}.
Several talks at this conference are devoted to 
this subject with different approaches. We refer to \cite{Panzer:LL2014,Panzer:2014fla} where Schwinger's $\alpha$ representation  of 
Feynman integrals 
is used for their evaluation 
with hyperlogarithms (package \verb|HyperInt|), and to the numerical approach \cite{Borowka:2014wla,Borowka:2014posLL2014} to 
the Feynman parameter 
representation with sector decomposition (package \verb|SecDec|).  
% Here, we follow another attractive method, relying on changing integrals over Feynman parameters into complex path 
% integrals using the MB identity.

Our starting point is the Feynman parameterization for a typical loop momentum integral of the form
\begin{eqnarray}
 G(X) & = &  
 \frac{(-1)^{N_{\nu}} \Gamma\left(N_{\nu}-\frac{d}{2}L\right)}{\prod \limits_{i=1}^{N}\Gamma(n_i)}
 \int \prod \limits_{j=1}^N dx_j ~ x_j^{n_j-1} \delta(1-\sum \limits_{i=1}^N x_i)
 \frac{U(x)^{N_{\nu}-d(L+1)/2}}{F(x)^{N_{\nu}-dL/2}}.
\label{basicFU}
\end{eqnarray} 
Here, $n_i$ is the power of inverse propagator $i$, $N_{\nu}=\sum_{i=1}^{N} n_i$, $d=4-2\epsilon$, $N$ the number of internal lines and 
$L$ the number of loops.
The functions $U$ and $F$ are called graph or Symanzik polynomials \cite{Nakanishi:1961,Nakanishi:1971,Bogner:2010kv} and are 
composed of the Feynman 
parameters $x_i$ ($U$ and $F$) and of the masses and momenta of the propagators ($F$).
The general MB relation can be applied to the polynomials~ $U$ and~ $F$:
\begin{eqnarray}
&        & \frac{1}{(A_1+\ldots+A_n)^{\lambda}} = \frac{1}{\Gamma(\lambda)}\frac{1}{(2 \pi i)^{n-1}} \nonumber \\ 
& \times & \int_{-i \infty}^{i \infty} dz_1 \ldots dz_{n-1} 
           \prod \limits_{i=1}^{n-1} A_i^{z_i} ~ A_n^{-\lambda - z_1 - \ldots - z_{n-1}}
           \times
           \prod \limits_{i=1}^{n-1} \Gamma(- z_i) ~
           \Gamma(\lambda + z_1 + \ldots + z_{n-1}).
\label{mbrel}
\end{eqnarray}
The result are multi-dimensional MB integrals.
   
The aim of our work is to make the procedure of solving these integrals automatic as much as possible.   
In general  we want  to  
\begin{enumerate}
   \item construct   \mb{} representations; 
   \item change them into nested sums; 
   \item solve the sums numerically or analytically. 
\end{enumerate}
   
Here we discuss shortly some aspects related to these steps.
Certainly, there are limitations for the application of the \mb{} method, whose complexity rises with
\begin{itemize}
  \item the number $L$ of the loops involved;
  \item the number of scales resulting from up to $N$ different internal masses and $E$ external massive and/or off-shell particles; 
  \item the number $N$ of internal lines;
  \item the rank $R$ in case of tensor integrals; $R=0$ for scalar integrals. 
\end{itemize}

\begin{figure}[t]
 \begin{center}
 \includegraphics[scale = 0.7]{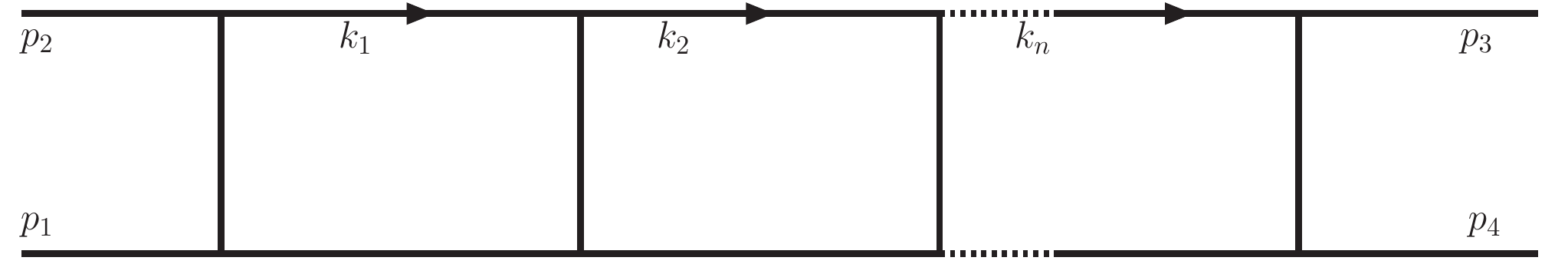}
 \end{center}
 \caption{\it $n$-loop ladder diagram with $k_1,...,k_n$ internal and $p_1,...,p_4$ external momenta.}
 \label{ladder}
\end{figure}

There are several public software packages for  the application of \mb{} integrals in particle physics calculations. 
On the MB Tools webpage \cite{mbtools} and on the \ar{}  webpage \cite{Katowice-CAS:2007} there are codes related to the MB approach:
 \begin{itemize}
\item \mb{} by M. Czakon \cite{Czakon:2005rk} and \verb|MBresolve| by V. Smirnov \cite{Smirnov:2009up} -- for the analytic continuation of 
Mellin-Barnes integrals in $\epsilon$;
\item  \verb|MBasymptotics| by M. Czakon (\cite{mbtools}, 2005) -- for the parametric expansions of Mellin-Barnes integrals;
\item  \verb|barnesroutines| by D. Kosower (\cite{mbtools}, 2008)  -- for the automatic application of the first and second Barnes lemmas;
\item  \ar v.2.0 \cite{Gluza:2007rt,Gluza:2010rn,Katowice-CAS:2007} -- for the creation of (mostly planar) MB representations and later 
calls of other tools;
\item \ar v.3.0  %np 
by I. Dubovyk, J. Gluza, K. Kajda, T. Riemann (\cite{Katowice-CAS:2007}, 2014) -- for the creation of 
non-planar MB representations up 
to two loops; the automatic three-loop case is under development. 
\end{itemize}
The \ar{} package generates multiloop scalar and tensor Feynman integrals. For planar cases, the  automatic derivation of \mb{} integrals 
by \ar{} is optimal when using the so-called loop-by-loop approach (\la{}). An example are the ladder diagrams shown in Fig.~\ref{ladder} 
and commented in Tab.~\ref{planartable}.

In this contribution, we present a new branch of the \ar{} package,  \ar v.3.0 (July 2014).
This version may treat also non-planar structures with improved efficiency and may treat up to  two loops presently. 
A global approach (\ga{}) may be chosen where $F$ and $U$ 
polynomials are changed into \mb{} representations with help of 
Eq.~(\ref{mbrel}) just in one step. 
Alternatively, one may choose a hybrid approach which treats planar subloops separately.
The procedures are discussed in section \ref{secmb}. 
For more details on the \la{} and \ga{} approaches, see \cite{Gluza:2007rt}.
 In section \ref{secsums} a few comments are given on how \mb{} integrals 
can be changed into nested sums, and how to calculate them. 
The paper finishes with a summary.

%\begin{widetext}
\begin{table}[b]
%\begin{center}
\begin{tabular}{|l|cccc|cccc|}
\hline \hline
Dimensions of                 & \multicolumn{4}{|c|}{Massless~ cases} & \multicolumn{4}{|c|}{Massive~ cases}  \\
planar ladder \mb{} integrals & \multicolumn{4}{|c|}{ }              & \multicolumn{4}{|c|}{ }              \\
\hline 
Number of loops ($L$) & 1 & 2 & 3 & 4  &      {1}          &  {2}          &  {3}           &  {4}              \\
\hline 
No Barnes First Lemma (BFL) & 1 & 4 & 7 & 10 &       3           &   8           &   13           &   18              \\
With BFL              &{1}&{4}&{7}&{10}&       2 ({1}+{1}) &   6 ({4}+{2}) &   10 ({7}+{3}) &   14 ({10}+{4})   \\
\hline \hline
\end{tabular}
\caption[]{\it Optimal results for ladder diagrams defined in Fig.~\ref{ladder}: $\rm{Dim(massive~ case)}=\rm{Dim(massless~ 
case)}+\#loops$.}
%}
%\end{center}
\label{planartable}
\end{table}
% \end{widetext}

%%%%%%%%%%%%%%%%%%%%%%%%%%%%%%%%%%%%%%%%%%%%%%%%%
\section{\label{secmb}Construction of Mellin-Barnes integrals}
%%%%%%%%%%%%%%%%%%%%%%%%%%%%%%%%%%%%%%%%%%%%%%%%%
For non-planar diagrams the \ga{} approach is applied. Further, we use the Cheng--Wu theorem \cite{Nakanishi:1971,SmirnovVA:2009} which 
states that Eq.~(\ref{basicFU}) holds also with the modified $\delta$-function
\begin{equation}
 \delta\left( \sum \limits_{i \in \Omega} x_i -1 \right),
\end{equation}
where $\Omega$ is an arbitrary subset of the lines $1, \ldots, L$, when the integration over
the rest of the variables, i.e. for $i \notin \Omega$, is extended to the integration from
zero to infinity. 
One can prove this theorem in a simple way 
starting from the $\alpha$-representation using
\begin{equation}
 1 = \int \limits_{0}^{\infty} \frac{d \lambda}{\lambda} 
 \delta \left( 1 - \frac{1}{\lambda} \sum \limits_{i=1}^{N} \alpha_i  \right) \Leftrightarrow
 1 = \int \limits_{0}^{\infty} \frac{d \lambda}{\lambda} 
 \delta \left( 1 - \frac{1}{\lambda} \sum \limits_{i \in \Omega} \alpha_i  \right)
\end{equation}
and changing variables from $\alpha_i$ to $\alpha_i = \lambda x_i$ as shown above.

%\section{Non--Planar Topology}          

Let us see how this works in the case of the non-planar massless  two-loop box, Fig.~\ref{figlnp}. This integral was first considered in 
\cite{Tausk:1999vh} and the  Symanzik polynomials are: 
\begin{figure}[t]
 \begin{center}
 \includegraphics[scale = 0.7]{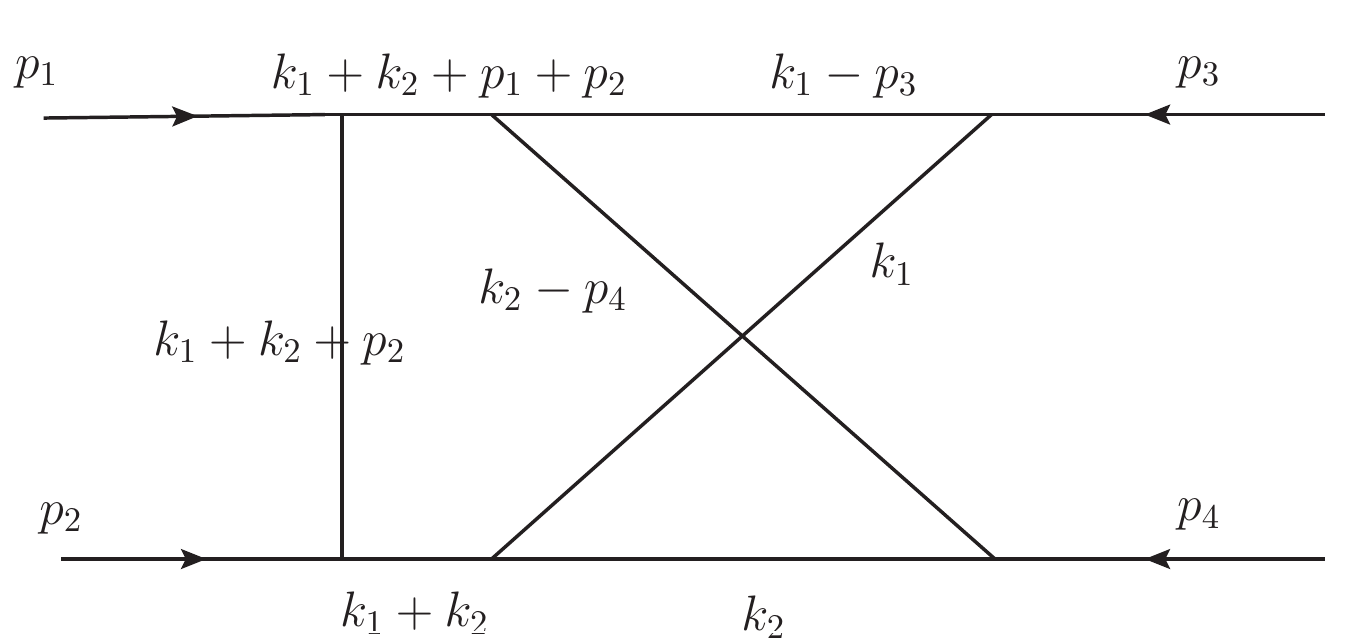}
 \end{center}
 \caption{\it The non-planar massless double box.}
 \label{figlnp}
\end{figure}
\begin{eqnarray}
  U(x) & = & (x_1 + x_6)(x_2 + x_7) + (x_3 + x_4 + x_5){(x_1 + x_2 + x_6 + x_7)}     ,        \\ %\nonumber
  F(x) & = & - t~ x_1 x_4 x_7 - u~ x_2 x_4 x_6 - s~ x_1 x_2 x_5 - s~ x_3 x_6 x_7   %\nonumber \\  &   &
 - s~ x_3 x_5 {(x_1 + x_2 + x_6 + x_7)}.
\end{eqnarray}
 
The factorizations in $U$ and $F$ are connected with a proper collection of Feynman parameters in 
front of the variables $k_1,k_2$, as shown schematically in Fig.~\ref{schemeF}.
Next we will apply the Cheng-Wu theorem. Then the integral 
becomes, after introduction of the $x$ variables: 
\begin{figure}[b]
 \begin{center}
 \includegraphics[width=\textwidth, height=3.5cm]{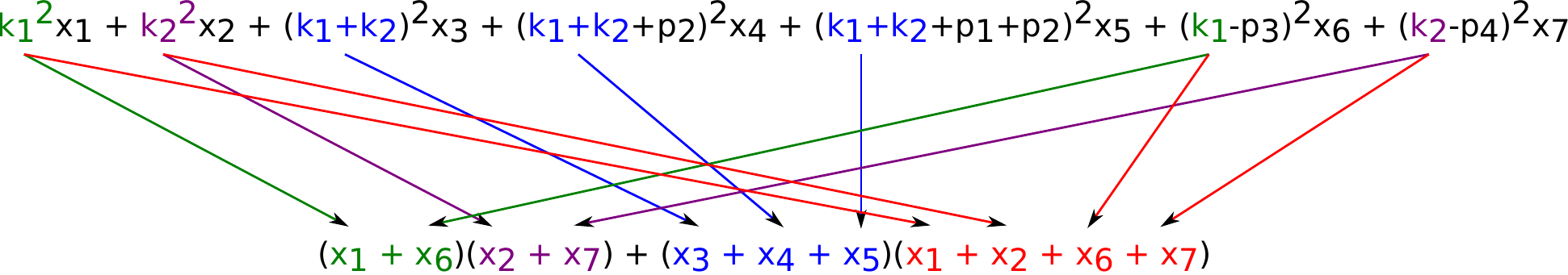}
 \end{center}
 \caption{\it Efficient factorization scheme for the $U$ polynomial.}
 \label{schemeF}
\end{figure}
\begin{eqnarray}
B_7^{NP} & = & \frac{(-1)^{N_{\nu}} \Gamma\left(N_{\nu}-d\right)}{\Gamma(n_1) \ldots \Gamma(n_7)}
               \int \limits_0^{\infty} dx_3 dx_4 dx_5 \int \limits_0^{1} dx_1 dx_2 dx_6 dx_7 \prod \limits_{j=1}^N  ~ x_j^{n_j-1}  \delta[1-{(x_1+x_2+x_6+x_7)}] \nonumber \\
         &   & \times~ \frac{((x_1 + x_6)(x_2 + x_7) + x_3 + x_4 + x_5)^{N_{\nu} - \frac{3d}{2}}}
               {(- t~ x_1 x_4 x_7 - u~ x_2 x_4 x_6 - s~ x_1 x_2 x_5 - s~ x_3 x_6 x_7 - s~ x_3 x_5)^{N_{\nu} - d}},
\end{eqnarray}
 and after resolving part of the $F(x)$ by Mellin-Barnes integrals:
\begin{eqnarray}
B_7^{NP} & = & \frac{(-1)^{N_{\nu}}}{\Gamma(n_1) \ldots \Gamma(n_7)} \int \limits_{-i \infty}^{i \infty} dz_1 \ldots dz_4 
               \int dx_1 \ldots dx_7 
               \prod \limits_{j=1}^N ~ x_j^{n_j-1} \delta[1-{(x_1+x_2+x_6+x_7)}] 
\nonumber \\ &   &\times~
               (-s)^{ - N_{\nu} + d - z_2 - z_3} (-t)^{z_2} (-u)^{z_3}            \nonumber \\
         &   & {\times~ \Gamma (-z_1) \Gamma (-z_2) \Gamma (-z_3) \Gamma (-z_4) 
               \Gamma(N_{\nu} - d + z_1 + z_2 + z_3 + z_4)}                       \nonumber \\
         &   & {\times~ x_1^{ - N_{\nu} + d - z_1 - z_2 - z_3} x_2^{z_2 + z_3}
               {x_3}^{ - N_{\nu} + d - z_2 - z_3 - z_4} {x_4}^{z_1 + z_3} {x_5}^{z_2 + z_4}   
               x_6^{z_1 + z_2} x_7^{z_3 + z_4} }                                  \nonumber \\
         &   & \times~ [{ x_3 + x_4 + x_5} + (x_1 + x_6)(x_2 + x_7)]^{N_{\nu} - \frac{3d}{2}}.
\end{eqnarray}
 Now we perform the integrations over Feynman parameters $x_i$.
In particular, $x_3,x_4,x_5$ can rather be called Cheng-Wu variables, for which we may use the Beta-function:
\begin{equation}
  \int \limits_0^{\infty} dx ~ x^{N_1}(x + A)^{N_2} = \frac{ A^{1 + N_1 + N_2} \Gamma(1 + N_1) \Gamma(-1 - N_1 - N_2)}{\Gamma(-N_2)}.
\end{equation}
This, and 
\begin{eqnarray}
\int \limits_0^{1} \prod_{i} dx_i x_i^{M_i-1}\delta(1-\sum_{i} x_i) = \frac{\prod_{i} \Gamma(M_i)}{\Gamma(\sum_{i} M_i)}
\end{eqnarray}
leads finally to the 4-dimensional Mellin-Barnes representation, corresponding, up to some shifts of variables, to Eq.~(8) of 
\cite{Tausk:1999vh}:
 \begin{eqnarray}
B_7^{NP} & = & \frac{(-1)^{N_{\nu}}}
               {\Gamma(n_1) \ldots \Gamma(n_7)} \int \limits_{-i \infty }^{i \infty} dz_1 \ldots dz_4
               (-s)^{4 - 2 \epsilon - N_{\nu} -z_{23}} (-t)^{z_3} (-u)^{z_2}                                %\nonumber 
\\
         &   &\times~ \frac{\Gamma(- z_1) \Gamma(- z_2) \Gamma(- z_3) \Gamma(- z_4) \Gamma(2 - \epsilon - n_{45}) 
               \Gamma(2 - \epsilon - n_{67})} {\Gamma(4 - 2 \epsilon -  n_{4567}) \Gamma(n_{45} +  z_{1234})
               \Gamma(n_{67} +  z_{1234}) \Gamma(6 -3 \epsilon -  N_{\nu})}                                 \nonumber \\
         &   &\times~ \Gamma(n_2 + z_{23}) \Gamma(n_4 + z_{24}) 
               \Gamma(n_5 + z_{13} ) \Gamma(n_6 + z_{34}) \Gamma(n_7 + z_{12})
               \Gamma^3(- 2 + \epsilon + n_{4567} + z_{1234})                                               \nonumber \\ 
         &   &\times~ \Gamma(4 - 2 \epsilon - n_{124567} - z_{123}) 
               \Gamma(4 -2 \epsilon -  n_{234567} -  z_{234}) 
               \Gamma(- 4 + 2 \epsilon +  N_{\nu} +  z_{1234}),\nonumber
\label{eq-mbtausk}
\end{eqnarray}
with notations $z_{i\ldots j \ldots k} = z_i + \ldots + z_j + \ldots + z_k$ and
$n_{i\ldots j \ldots k} = n_i + \ldots + n_j + \ldots + n_k$. 
 
In the new version of the \ar{} package, \ar v.3.0 \cite{ambrenp10:2014},
the procedure is applied properly in an 
automatic way. 
The corresponding Mathematica sample file \texttt{MB\_\-GA\_\-2loop\-NP\_\-massless.nb} with a derivation of Eq.~(\ref{eq-mbtausk}) can be 
found at \cite{Katowice-CAS:2007}.
%\href{http://prac.us.edu.pl/~gluza/ambre/}{http://prac.us.edu.pl/~gluza/ambre/}.
%\url{http://prac.us.edu.pl/~gluza/ambre/}.  %{http://prac.us.edu.pl/~gluza/ambre/}.
% More details will be described in a forthcoming paper. 
 
\bigskip

\begin{figure}
 \begin{center}
 \includegraphics[scale = 0.7]{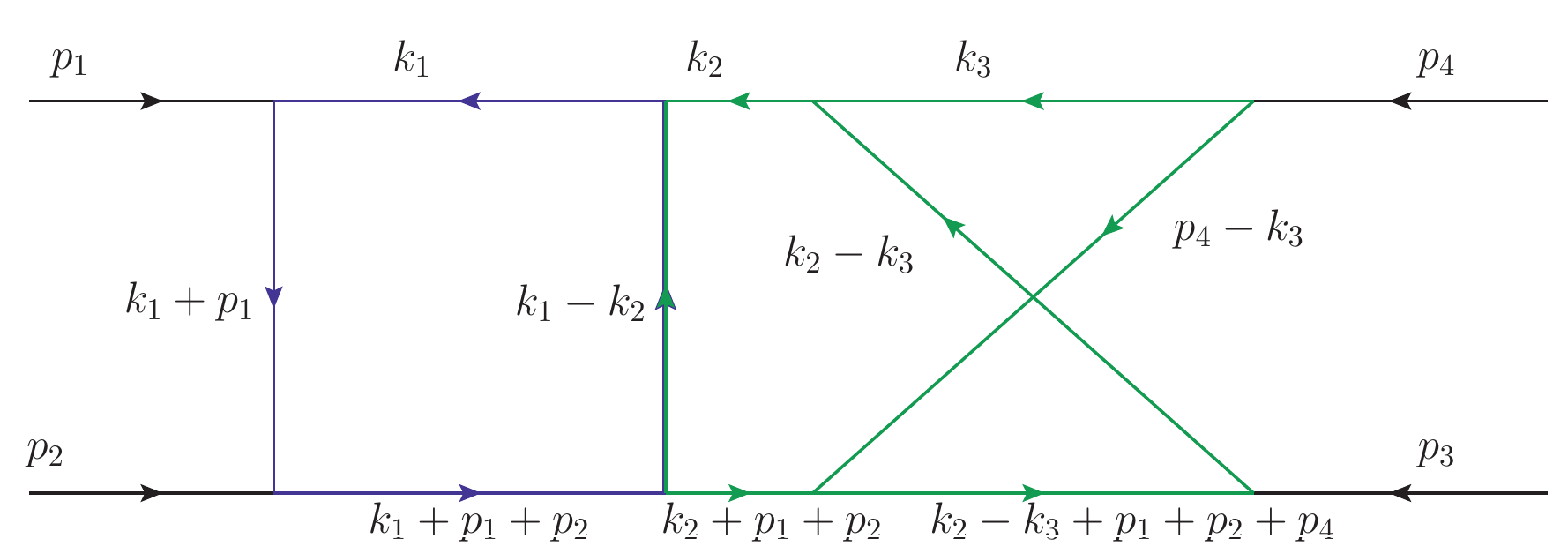}
 \caption{\it Hybrid method: first $F$ and $U$ polynomials over the ${k_1}$-subloop are worked out (planar subgraph), 
          then $\{k_2, k_3\}$ are integrated over (non-planar subgraph).}
 \label{figlhyb}
 \end{center}
\end{figure}

In the same way we can use the Cheng-Wu theorem for massless non-planar 
3-loop diagrams, e.g. for that shown in Fig.~\ref{figlhyb}. Here we prefer to use a hybrid method. 
In a first stage, the planar subloop over $k_1$ is considered and the 
corresponding $F$ and $U$ polynomials are changed into \mb{} integrals. 
Next the non-planar subloop with momenta  $k_2,k_3$ is considered 
using the \ga{} approach for non-planar diagrams. 
For the planarity identification, i.e. the check whether a diagram is planar or not, 
the Mathematica package \verb|PlanarityTest| v.1.1 \cite{planaritytest:2014}  is used 
\cite{Bielas:2013rja}. 
%http://prac.us.edu.pl/~gluza/ambre/planarity/
An example for the hybrid method can be found at the webpage \cite{Katowice-CAS:2007}, as file \texttt{MB\_hybrid\_3loopNP\_massless.nb}.
% An example for the hybrid method can be 
% found at \url{http://prac.us.edu.pl/~gluza/ambre/}, as file \verb|MB_hybrid_3loopNP_massless.nb|.

In Tables~\ref{table2} and \ref{table3}, the numbers of dimensions of the Mellin-Barnes representation for the massive 3-loop cases are 
shown for the loop-by-loop approach (\la) and for the general approach (\ga), respectively.
An efficient and satisfactory way to decrease them has not been found so far.  
More details on the new implementations in the \ar{} package will be described in a forthcoming paper.

\begin{table}[b]
\begin{center}
\setlength{\tabcolsep}{4pt}
\begin{tabular}{rr|rr}
\hline \hline
\multicolumn{2}{l}{Massless~ case} & \multicolumn{2}{l}{Massive~ case} \\
\hline \hline
 2-loop   & 3-loop &   2-loop   & 3-loop     \\
   7      &   10   &     11     &   16       \\
   6      &   9    &     8      &   12       \\
\hline \hline
\end{tabular}
\end{center}
%\caption{}
\caption[]{\it Mellin-Barnes integrals derived by \ar{}. The loop-by-loop approach (\la) is applied: dimensions of some $2\rightarrow2$ 
{non--planar} 
topologies before and after applying Barnes' first
Lemma. The 2-loop topology corresponds to Fig.~\ref{figlnp}, while the 3-loop topology corresponds to Fig.~\ref{figlhyb}. 
Massive case means that the horizontal lines in Figs.~\ref{figlnp} and ~\ref{figlhyb} are massive.}
\label{table2}
\end{table}

\begin{table}[t]
\setlength{\tabcolsep}{4pt}
\begin{center}
\begin{tabular}{rr|rr}
\hline \hline
\multicolumn{2}{l}{Massless case} & \multicolumn{2}{l}{Massive~ case} \\
\hline \hline
 2-loop    & 3-loop &   2-loop  & 3-loop     \\
     4    &   7   &      13    &  18         \\
     4    &   7   &      11    &  15         \\
\hline \hline
\end{tabular}
\end{center}
\caption[]{\it Mellin-Barnes integrals derived by \ar{}. The general approach (\ga) is applied. All notations are as in Table 
\ref{table2}.}
\label{table3}
\end{table}

%%%%%%%%%%%%%%%%%%%%%%%%%%%%%%%%%%%%%%%%%%%%%%%%%%%
\section{\label{secsums}Towards summation of iterated integrals}
%%%%%%%%%%%%%%%%%%%%%%%%%%%%%%%%%%%%%%%%%%%%%%%%%%%%%%
In a second part of our project of developing automatic tools for deriving and solving Mellin-Barnes integrals  
we explore the general case of deriving multiple sums of residues for Feynman integrals, with the idea then to apply algebraic summation 
techniques like those developed at the Research Institute for Symbolic Computation (RISC) ~\cite{Riemann:2012linz}.
As a first step we prepared an appropriate Mathematica package, \verb|MBsums| v.0.9 \cite{mbsums:2014}.
%(author: M. Ochman, July 2014)}.

In the approach to Feynman integrals advocated here, the \mb{} representations  can be cast in a 
general form:
 
\begin{equation}
\frac{1}{(2\pi i)^r}\int\limits_{-i \infty}^{i \infty} \dots \int\limits_{-i \infty}^{i \infty} \underset{i}{\overset{r}{\Pi}} dz_i \;
    {{F}}(Z,X,{\vec{n}},\epsilon)  \frac{\underset{j}{\Pi}\;
    {{G_j}}(N_j)}{\underset{k}{\Pi}\; {{G_k}}(N_k)} .
    \label{mbgen}
\end{equation}
The integral ranges over $r$ complex variables $z_i \in Z$, and the integration path has to be properly chosen.
The  function $F$ depends on kinematical parameters like $s,t,u$, which are derived from the scalar invariants of the problem,  and on  
internal masses; they are represented by $X$.
The  indices of the propagators $n_i$ constitute  ${\vec n} = \{n_1,\ldots, n_N\}$, and the dimensional regulator is $\epsilon=2-D/2$.

In practice $F$ is a   product of powers of dimensionless parameters $X_k$, with exponents being linear   combinations of $z_i,n_i$ and 
$\epsilon$:
   \begin{equation}
 {F} \sim \prod_{k} \; {X}_k^{\sum_{i,j} ( \alpha_i z_i + \beta_j n_j + \gamma \epsilon)},
  \end{equation}
where the coefficients  $\alpha_i,\beta_j,\gamma$ are real,  $\alpha_i,\beta_j,\gamma \in \mathbf{R}$, and  the ${X_k}$ 
are composed of ratios of invariants and masses, e.g.
  ${X_k} \in \{\frac{ s}{ t},\frac{ m^2}{ s},...\}.$
The $\Gamma$-functions $G_k(N_k)$ in Eq.~(\ref{mbgen}) have arguments similar to the exponents occurring in $F$, 
$N_k = \sum_{i,j} ( \alpha_i z_i + \beta_j n_j + \delta \epsilon)$.   

The point is how to evaluate the integral in  Eq.~(\ref{mbgen}) exactly or in some approximation, analytically or numerically?
Assuming the kinematics to be euclidean or minkowskian, allowing for arbitrary numerics or assuming some small parameters in the set $X$ 
(see e.g. \cite{Czakon:2006pa}).

A combined use of \ar{} and \mb{} allows to derive a collection of Mellin-Barnes representations for scalar or tensor Feynman integrals, 
already regulated in the dimension $D=4-2\epsilon$.

Least problems are faced when seeking some numerical answers for euclidean kinematics. A combined use of \ar{} and \mb{} with a numerical 
integration package often is applicable.
For massless problems, one or the other more or less systematic approach towards analytical solutions is known 
\cite{Moch:2005uc,Brown:2008um,Anastasiou:2013srw,Panzer:LL2014}.

For massive, non-planar Feynman integrals however, it is nearly impossible to find  complete analytic solutions.
Only one of the problems is that \mb{} integrals start to be truly multidimensional. 

An immediate method is changing the regulated  \mb{} integrals into (usually) infinite series by {closing the  integration contours in the 
multi-dimensional complex plane and calculating the sequence of residues.
Depending on closing contours to the left or right, convergent infinite series can be obtained which might be suitable for further 
processing. 
However, the choice of proper contours depends not only on the general features of the Feynman integral, but also on the values of 
kinematical parameters.
Let us look at the simple one-dimensional integral 
\begin{equation}\label{massive-qed-vertex-constant}
\int\limits_{-i \infty-1/2}^{i \infty-1/2} dz_1(-1/s)^{z_1} \frac{\Gamma[-z_1]^3 \Gamma[1+z_1]}{\Gamma[-2 z_1]}
\end{equation}
which corresponds (up to a factor) to the coefficient of the $1/\epsilon$ term in the expansion of the massive QED box function.
It might look convergent for e.g. $s=-3$ when closing the $z_1$ contour to the right and taking residues, but in 
fact the Gamma functions of the integrand change the sum into a chain of alternating and increasing numbers. The integral  
converges above the threshold,  $|s|>4$, if the contour is closed to the right.

The input function of the Mathematica package \verb|MBsums| is one of the two following:
\begin{equation} \label{exampleMBgen} 
\verb|MBIntToSum[int,{},contours]|,  
\end{equation}
\begin{equation} \label{exampleMBgen2} 
\verb|MBIntToSum[int,kinematics,contours]|.
\end{equation}
Here \verb|int| is the expression to be evaluated (an output from \ar{} and \mb{}, with or without a dependence on $\epsilon$). 
The list \verb|contours| is connected with choices of (i) the order of integrations, and (ii) closing integration contours (to the left (L) 
or to the right (R)). 
If the list of kinematics is not empty, the list of contours will be automatically changed by  \verb|MBsums| in order to achieve 
numerical convergence at the specified kinematical point. 
% Without specific 
% choice, it is calculated in an automatic way, taking into account a specific numerical value of a kinematic point.
For instance, 
for a two dimensional \mb{} integral which emerges in some massive cases, we have:
%%%%%%%%%%%%% old 
% \begin{equation}\label{exampleMB} 
%  \verb| int = MBint[((-s/m^2)^(z1 - z2) Gamma[1 - z1] Gamma[-z1] |
% \end{equation}
%       \begin{verbatim}
%       Gamma[z1] Gamma[1 + z1] Gamma[1 + z1 - z2]^2 
%       Gamma[-z2] Gamma[-z1 + z2])/Gamma[2 + z1 - 2 z2], 
%       {{eps -> 0}, {z1 -> -(13/32), z2 -> -(5/32)}}],
%           \end{verbatim}
\begin{eqnarray}\label{exampleMB}
&&\verb|int = MBint[((-x)^(z1 - z2) Gamma[1 - z1] Gamma[-z1] |\\
&&\verb|Gamma[z1] Gamma[1 + z1] Gamma[1 + z1 - z2]^2|\nonumber\\
&&\verb|Gamma[-z2] Gamma[-z1 + z2])/Gamma[2 + z1 - 2 z2],|\nonumber\\
&&\verb|{{eps -> 0}, {z1 -> -(13/32), z2 -> -(5/32)}}]|\nonumber .
\end{eqnarray}
The corresponding function call might be:
\begin{equation}\label{exampleMB2} 
   \verb|MBIntToSum[int,{x->-1/2},{z2 -> L, z1 -> L}]|.
\end{equation}
In this case the output is:
\begin{eqnarray}\label{exampleMB3}
&&\verb|{{((-1)^(-n1 - 2 n2) (-x)^n1 n1! (-1 + n1 + n2)!|
\\
&&\verb|(HarmonicNumber[-1 + n1 + n2]- HarmonicNumber[1 + 2 n1 + n2]))|
\nonumber\\
&&\verb|/(1 + 2 n1 + n2)!, n1 >= 0 && n2 >= 1, {n1, n2}}}|.
\nonumber
\end{eqnarray}
% \begin{eqnarray}\label{exampleMB3}
% &&\verb|{(x^n2 n2! (n1 + n2)! (HarmonicNumber[n1 + n2] |\\
% &&\verb|- HarmonicNumber[2 + n1 + 2 n2]))/(2 + n1 + 2 n2)!,|\nonumber\\
% &&\verb|n1 >= 0 && n2 >= 0}|\nonumber
% \end{eqnarray}
% \begin{equation}\label{exampleMB3} 
% \verb|{(B^n2 n2! (n1 + n2)! (HarmonicNumber[n1 + n2] |
% \end{equation}
% \begin{verbatim}
%     - HarmonicNumber[2 + n1 + 2 n2]))/(2 + n1 + 2 n2)!,
%       n1 >= 0 && n2 >= 0}
% \end{verbatim}
This expression represents a double sum in $x=-s/m^2$, %$B=-s/m^2$, 
to be summed over non-negative \texttt{n1} and \texttt{n2>0}.
For details see also the sample file \verb|LL2014-MBsums.nb| at \cite{mbsums:2014}.

Here the situation is relatively simple, but more general conditions are possible, e.g.:  \texttt{n1 >= 0 \&\& n2 >= 0 \&\& n1 > n2}.
Certainly, for higher dimensional integrals, the multiple series will be much more involved, as far as conditional statements for the  
counting parameters $n_i$ are concerned. For a $D$-dimensional multiple integral, the list  \verb|contours| can be given in $D! 2^D$ 
different ways, depending on the integration order and on the direction of closing the contours. In the above example, there are eight 
possible versions. One may try them out and will observe final expressions of rather different complexity; the most efficient one has been 
shown here.
We have no recipe to optimize the program properly in this respect and leave this to the user.
Another tricky point for programming was the proper taking into account of multiple residues arising from singularities of several Gamma 
functions (and their derivatives) at 
certain values of the summation variables. 

After getting proper multiple series, one may proceed with their summation, %with one method or the other.
using advanced summation technologies
\cite{Karr:81,Schneider:01,Schneider:05a,Schneider:07d,Schneider:08c,Schneider:10a,Schneider:10b,Schneider:10c,
Schneider:13b} as encoded in the packages {\tt Sigma} \cite{SIG1,SIG2}, {\tt EvaluateMultiSums}, and {\tt 
SumProduction} \cite{Ablinger:2010pb,Blumlein:2012hg,Schneider:2013zna}.

In fact, compared to the derivation of multiple series, which is at least in principle straightforward, the summation is the 
much more 
involved part of the problem. 
In an ideal world, one has a summation theory at hand which has been solving the summation problem already in generality, and which is made 
applicable in a computer algebra package.

In our real world, usually we do not have such a theory and/or package.

There are simple cases with known solutions.
One of them is the integral in Eq.~(\ref{massive-qed-vertex-constant}).
The corresponding infinite sum may be done just by Mathematica's built-in function \verb|Sum|, but also with 
available packages \cite{Gluza:2007bd,Huber:2005yg,Huber:2007dx,Davydychev:2000na}.

In our study of planar massive 2-loop box integrals for Bhabha scattering \cite{Czakon:2006pa}, we could sum up all the infinite 
sums by XSUMMER \cite{Moch:2005uc} after expanding in the small parameter $m_e^2/s$; the problem became low-dimensional (in fact: 
factorized 
into one-dimensional sums) and was solved by simple polylogarithmic functions.
In the famous cases of solving the planar and non-planar massless QED double boxes, as well as the planar massive one \cite{Smirnov:2001cm}, 
the authors were able to sum up their expressions until the constant term by dedicated handlings. For the non-planar massive double box, 
this was already impossible \cite{Heinrich:2004iq}.

We do not discuss here some additional numerical treatments.
One can try:
\begin{enumerate}
   \item Write the sum in terms of nested sums using the Mathematica package \texttt{Sigma} \cite{Schneider:2013zna}. 
   \item Exploit the structure of the sums by recursively applying rewrite rules to generate iterated integrals; see the talk by C. 
Raab  at this conference, \cite{Raab:LL2014,Ablinger:2014bra}.
   \item Simplify the result to a canonical form, e.g. by using the Mathematica package \\ \texttt{HarmonicSums} 
\cite{Ablinger:2010kw,Ablinger:2011te,Ablinger:2013cf,Ablinger:2013hcp}.
   \end{enumerate} 
% from JB 2014-07-10 TORD.tex
More specifically,  the aim is to rewrite infinite sums in terms of iterated integrals
\begin{equation}
     \sum_{n=0}^\infty f(n)x^n \quad\rightarrow\quad h_0(x)\int_0^xdt_1h_1(t_1)\dots\int_0^{t_{k-1}}dt_kh_k(t_k)
  \end{equation}
    over a certain alphabet of integrands, which occur in course of a systematic conversion, 
as e.g. 
   \begin{equation}\label{eq-alphabet}
\left\{     \frac{1}{t-a},\quad \frac{1}{(t-a)\sqrt{t-b}},\quad \frac{1}{\sqrt{t-b}\sqrt{t-c}},\quad \frac{1}{(t-a)\sqrt{t-b}\sqrt{t-c}}  
\right\},
 \end{equation}
with $a, b, c \in \mathbb{R}$.

For instance, in the case of the \mb{} integral in Eq.~(\ref{exampleMB}), we have

\begin{equation}
  \sum_{n_2=0}^\infty\sum_{n_1=0}^\infty\frac{n_2!(n_1+n_2)!}{(n_1+2n_2+2)!}
\left(S_1(n_1+n_2)-S_1(n_1+2n_2+2)\right)x^{n_2},
\end{equation}
with $S_1 (N)= \sum_{k=1}^N (1/k)$ the harmonic sum.

So, the steps to be applied are:

 \begin{enumerate}
  \item Compute the inner sum using \texttt{Sigma} \cite{Schneider:2013zna}:
    \begin{equation}
      \sum_{n_2=0}^\infty\frac{S_1(n_2)-S_1(2n_2+1)}{(n_2+1)(2n_2+1)\binom{2n_2}{n_2}}x^{n_2} .
    \end{equation}
  \item Transform the result algorithmically to iterated integrals 
   \begin{equation} 
\frac{1}{x}\int_0^x\frac{dt_1}{\sqrt{t_1}\sqrt{4-t_1}}\int_0^{t_1}\frac{dt_2}{\sqrt{t_2}\sqrt{4-t_2}}\left(-2+\int_0^{t_2}\frac{dt_3}{t_3^{ 
3/2}\sqrt{4-t_3}}\int_0^{t_3}\frac{t_4dt_4}{\sqrt{t_4}\sqrt{4-t_4}}\right) .
   \end{equation}
  \item Convert the integrals to normal form using integrands of the form (\ref{eq-alphabet}) only, %Compactify the expressions
   \begin{equation}
    -\frac{1}{x}\int_0^x\frac{dt_1}{\sqrt{t_1}\sqrt{4-t_1}}\int_0^{t_1}\frac{dt_2}{t_2}\int_0^{t_2}\frac{dt_3}{\sqrt{t_3}\sqrt{4-t_3}}.
   \end{equation}
  \item In the present simple case one may
remove square roots by the transformation $x=-\frac{(1-y)^2}{y}$, see e.g. \cite{Davydychev:2003mv}, 
rewrite the result by \texttt{HarmonicSums} 
\cite{Ablinger:2010kw,Ablinger:2011te,Ablinger:2013cf,Ablinger:2013hcp},
and express the final harmonic polylogarithms \cite{Remiddi:1999ew} in terms of the classical polylogarithms
   \begin{equation}
    \frac{y}{(1-y)^2}\left(4\zeta_3+2\zeta_2\ln(y)+\frac{\ln(y)^3}{6}+2\ln(y)\mathrm{Li}_2(y)-4\mathrm{Li}_3(y)\right).
  \end{equation}
 \end{enumerate}
In general a much wider class of iterated integrals will occur, see e.g. 
\cite{Ablinger:2014bra,Ablinger:2013jta}.

We are far from a stage where we might say that a summation procedure exists or where an automatic procedure would apply.
It is even unknown in which kind of function space the result of a multiple sum might be found. 

But we see from the more developed method of differential equations for solving Feynman integrals that the research field is promising.

%%%%%%%%%%%%%%%%%%%%%%%%%%%%%%%%%%%%%%%%%%%%%%%%%%%%%%%%%%%%%%%%%%%%%%%%%%%%%%%%%%%
\section{\label{secsummary}Summary}
The \mb{} approach for the calculation of Feynman diagrams relies on an individual approach to single multi-dimensional 
complex contour integrals. There are tools which help to solve them, though still not general enough for the massive cases we are 
interested 
in. 
We reported here on some progress in constructing \mb{} representations with a new version of the \ar{} package, being better suited for 
non-planar Feynman diagrams, and on progress in changing them into suitable multiple sums, which might be further studied aiming at 
analytic 
solutions. At the moment the main concern in our investigations focuses on systematic solutions of two- and three-dimensional planar 
and non-planar \mb{} integrals corresponding to massive Feynman integrals.
Using a specific, relatively simple case of a double sum arising from a massless configuration, we sketched the subsequent algebraic 
treatment with algorithms of summation theory.   
\section*{Acknowledgements}
We thank Bas Tausk for helpful discussions on non-planar Feynman integrals. \\
 Work supported by the Research Executive Agency (REA) of the European 
 Union under the Grant Agreement numbers PITN-GA-2010-264564 (LHCPhenoNet) and 
PITN-GA-2012-316704 ({HIGGS\-TOOLS}),
the Polish National Center of Science (NCN) under the Grant Agreement
number DEC-2013/11/B/ST2/04023,
by DFG Sonderforschungsbereich Transregio 9, Computergest{\"u}tzte Theoretische Teilchenphysik, 
and the Austrian Science Fund (FWF) grants P20347-N18 and SFB F50 (F5009-N15).

% \begin{align}\label{exampleMB3}
% \begin{verbatim}
% {(B^n2 n2! (n1 + n2)! (HarmonicNumber[n1 + n2] 
%     - HarmonicNumber[2 + n1 + 2 n2]))/(2 + n1 + 2 n2)!,
%       n1 >= 0 && n2 >= 0}
% \end{verbatim}
% \end{align}
% 
% \begin{Verbatim}[numbers=right,stepnumber=2,firstnumber=8]
% A=B
% \\
% 
% C+o
% \end{Verbatim}

% \input{gluza-ll14.bbl}
% \end{document}

\providecommand{\href}[2]{#2}
\addcontentsline{toc}{section}{References}

% \bibliographystyle{JHEP}
% %\bibliographystyle{epj} used by Ievgen
% %\bibliographystyle{elsarticle-num} used by Tord for Januzs' slides
% \bibliography{2loops}

\begin{thebibliography}{10}

\bibitem{zbMATH02640947}
E.~W. Barnes, A new development of the theory of the hypergeometric functions,
  Proc. Lond. Math. Soc. (2) 6 (1908) 141--177,
  \href{http://plms.oxfordjournals.org/content/s2-6/1/141.full.pdf}{http://plms.oxfordjournals.org/content/s2-6/1/141.full.pdf}.

\bibitem{Usyukina:1975yg}
N.~Usyukina, {\it {On a Representation for Three Point Function}},  {\em
  Teor.Mat.Fiz.} {\bf 22} (1975) 300--306,
  \href{http://dx.doi.org/10.1007/BF01037795}{DOI: 10.1007/BF01037795}.

\bibitem{Boos:1990rg}
E.~Boos and A.~I. Davydychev, {\it {A Method of evaluating massive Feynman
  integrals}},  {\em Theor.Math.Phys.} {\bf 89} (1991) 1052--1063,
  \href{http://dx.doi.org/10.1007/BF01016805}{DOI: 10.1007/BF01016805}.

\bibitem{Smirnov:1999gc}
V.~Smirnov, {\it {Analytical result for dimensionally regularized massless
  on-shell double box}},  {\em Phys. Lett.} {\bf B460} (1999) 397--404,
  [\href{http://xxx.lanl.gov/abs/hep-ph/9905323}{{\tt hep-ph/9905323}}].

\bibitem{Tausk:1999vh}
B.~Tausk, {\it {Non-planar massless two-loop {Feynman} diagrams with four on-
  shell legs}},  {\em Phys. Lett.} {\bf B469} (1999) 225--234,
  [\href{http://xxx.lanl.gov/abs/hep-ph/9909506}{{\tt hep-ph/9909506}}].

\bibitem{mbtools}
MB Tools webpage
  \href{http://projects.hepforge.org/mbtools/}{http://projects.hepforge.org/mbtools/}.

\bibitem{Czakon:2005rk}
M.~Czakon, {\it {Automatized analytic continuation of Mellin-Barnes
  integrals}},  {\em Comput. Phys. Commun.} {\bf 175} (2006) 559--571,
  [\href{http://xxx.lanl.gov/abs/hep-ph/0511200}{{\tt hep-ph/0511200}}].

\bibitem{Smirnov:2009up}
A.~V. Smirnov and V.~A. Smirnov, {\it {On the Resolution of Singularities of
  Multiple Mellin- Barnes Integrals}},  {\em Eur. Phys. J.} {\bf C62} (2009)
  445, [\href{http://xxx.lanl.gov/abs/0901.0386}{{\tt 0901.0386}}].

\bibitem{Gluza:2007rt}
J.~Gluza, K.~Kajda, and T.~Riemann, {\it {AMBRE - a Mathematica package for the
  construction of Mellin-Barnes representations for Feynman integrals}},  {\em
  Comput. Phys. Commun.} {\bf 177} (2007) 879--893,
  [\href{http://xxx.lanl.gov/abs/0704.2423}{{\tt 0704.2423}}].

\bibitem{Gluza:2010rn}
J.~Gluza, K.~Kajda, T.~Riemann, and V.~Yundin, {\it {Numerical Evaluation of
  Tensor Feynman Integrals in Euclidean Kinematics}},  {\em Eur. Phys. J.} {\bf
  C71} (2011) 1516, [\href{http://xxx.lanl.gov/abs/1010.1667}{{\tt
  1010.1667}}].

\bibitem{Panzer:LL2014}
E. Panzer, {\it Feynman integrals via hyperlogarithms}, talk held at Workshop
  {\it Loops and Legs in Quantum Field Theory}, April 27 - May 2, 2014, Weimar,
  Germany [\href{http://arxiv.org/abs/arXiv:1407.0074}{arXiv:1407.0074}];
  slides of talk at
  \href{https://indico.desy.de/conferenceDisplay.py?confId=8107}{LL2014
  webpage}.

\bibitem{Panzer:2014fla}
E.~Panzer, {\it {Feynman integrals via hyperlogarithms}},
  \href{http://xxx.lanl.gov/abs/1407.0074}{{\tt 1407.0074}}. For the package
  \verb|HyperInt| see \cite{panzer-hyperint}.

\bibitem{Borowka:2014wla}
S.~Borowka, T.~Hahn, S.~Heinemeyer, G.~Heinrich, and W.~Hollik, {\it
  {Momentum-dependent two-loop QCD corrections to the neutral Higgs-boson
  masses in the MSSM}},  \href{http://xxx.lanl.gov/abs/1404.7074}{{\tt
  1404.7074}}. Sector decomposition package \verb|SecDec| v.2.2,
  \href{http://secdec.hepforge.org}{http://secdec.hepforge.org}.

\bibitem{Borowka:2014posLL2014}
S.~Borowka, {\it {Momentum-dependent two-loop QCD corrections to the neutral
  Higgs-boson masses in the MSSM}},  {\em PoS} {\bf LL2014} (2014) 033. Slides
  of talk at
  \href{https://indico.desy.de/conferenceDisplay.py?confId=8107}{LL2014
  webpage}.

\bibitem{Nakanishi:1961}
N.~Nakanishi, {\it Parametric integral formulas and analytic properties in
  perturbation theory},  {\em Prog. Theor. Phys. Supplement} {\bf 18} (1961) 1,
  \href{http://ptps.oxfordjournals.org/content/18/1.full.pdf}{http://ptps.oxfordjournals.org/content/18/1.full.pdf}.

\bibitem{Nakanishi:1971}
N.~Nakanishi, {\em {Graph Theory and {Feynman} Integrals}}.
\newblock Gordon and Breach, 1971.

\bibitem{Bogner:2010kv}
C.~Bogner and S.~Weinzierl, {\it {Feynman graph polynomials}},  {\em Int. J.
  Mod. Phys.} {\bf A25} (2010) 2585--2618,
  [\href{http://xxx.lanl.gov/abs/1002.3458}{{\tt 1002.3458}}].

\bibitem{Katowice-CAS:2007}
I. Dubovyk, J. Gluza, K. Kajda, T. Riemann, \verb|AMBRE| project webpage,
  Silesian Univ., Katowice,
  \href{http://www.us.edu.pl/~gluza/ambre}{http://www.us.edu.pl/~gluza/ambre}.

\bibitem{SmirnovVA:2009}
V. Smirnov, {\it Multiloop Feynman integrals}, Scholarpedia, 4(6):8507,
  \href{http://dx.doi.org/10.4249/scholarpedia.8507}{http://dx.doi.org/10.4249/scholarpedia.8507}
  (June 23, 2014).

\bibitem{ambrenp10:2014}
I. Dubovyk, J. Gluza, K. Kajda, T. Riemann, Mathematica package \verb|AMBRE|
  v.3.0, webpage
  \href{http://www.us.edu.pl/~gluza/ambre/}{http://www.us.edu.pl/~gluza/ambre/}.

\bibitem{planaritytest:2014}
I. Dubovyk and K. Bielas, Mathematica package \verb|PlanarityTest|, webpage
  \href{http://www.us.edu.pl/~gluza/ambre/planarity}{http://www.us.edu.pl/~gluza/ambre/planarity}.

\bibitem{Bielas:2013rja}
K.~Bielas, I.~Dubovyk, J.~Gluza, and T.~Riemann, {\it {Some Remarks on
  Non-planar Feynman Diagrams}},  {\em Acta Phys. Polon.} {\bf B44} (2013),
  no.~11 2249--2255, [\href{http://xxx.lanl.gov/abs/1312.5603}{{\tt
  1312.5603}}].

\bibitem{Riemann:2012linz}
J. Gluza and T. Riemann, {\it Simple Feynman diagrams and simple sums}, Talk
  held at RISC - DESY {\it Workshop on Advanced Summation Techniques and their
  Applications in Quantum Field Theory} on the occasion of the 5th year jubilee
  of the RISC-DESY cooperation, May 7-8, 2012, RISC Institute, Castle of
  Hagenberg, Linz, Austria. Conference link:
  \href{http://www.risc.jku.at/conferences/RISCDESY12/}{http://www.risc.jku.at/conferences/RISCDESY12/},
  talk link:
  
\href{http://www-zeuthen.desy.de/~riemann/Talks/riemann-Linz-2012-05.pdf}{http://www-zeuthen.desy.de/~riemann/Talks/riemann-Linz-2012-05.pdf
  }.

\bibitem{mbsums:2014}
M. Ochman, Mathematica package \verb|MBsums|, webpage
  \href{http://www.us.edu.pl/~gluza/ambre/MBsums}{http://www.us.edu.pl/~gluza/ambre/MBsums}.

\bibitem{Czakon:2006pa}
M.~Czakon, J.~Gluza, and T.~Riemann, {\it {The planar four-point master
  integrals for massive two-loop Bhabha scattering}},  {\em Nucl. Phys.} {\bf
  B751} (2006) 1--17, [\href{http://xxx.lanl.gov/abs/hep-ph/0604101}{{\tt
  hep-ph/0604101}}].

\bibitem{Moch:2005uc}
S.~Moch and P.~Uwer, {\it {XSummer: Transcendental functions and symbolic
  summation in Form}},  {\em Comput. Phys. Commun.} {\bf 174} (2006) 759--770,
  [\href{http://xxx.lanl.gov/abs/math-ph/0508008}{{\tt math-ph/0508008}}].

\bibitem{Brown:2008um}
F.~Brown, {\it {The massless higher-loop two-point function}},  {\em Commun.
  Math. Phys.} {\bf 287} (2009) 925--958,
  [\href{http://xxx.lanl.gov/abs/0804.1660}{{\tt 0804.1660}}].

\bibitem{Anastasiou:2013srw}
C.~Anastasiou, C.~Duhr, F.~Dulat, and B.~Mistlberger, {\it {Soft triple-real
  radiation for Higgs production at N3LO}},  {\em JHEP} {\bf 1307} (2013) 003,
  [\href{http://xxx.lanl.gov/abs/1302.4379}{{\tt 1302.4379}}].

\bibitem{Karr:81}
M.~Karr, {\it Summation in finite terms},  {\em J.~ACM} {\bf 28} (1981)
  305--350.

\bibitem{Schneider:01}
C.~Schneider, {\it Symbolic summation in difference fields},  Tech. Rep. 01-17,
  RISC-Linz, J.~Kepler University, November, 2001.
\newblock PhD Thesis,
  \href{http://www.risc.jku.at/publications/download/risc_3017/SymbSumTHESIS.pdf}{
  http://www.risc.jku.at/publications/download/risc\_3017/SymbSumTHESIS.pdf}.

\bibitem{Schneider:05a}
C.~Schneider, {\it Solving parameterized linear difference equations in terms
  of indefinite nested sums and products},  {\em J. Differ. Equations Appl.}
  {\bf 11} (2005) 799--821,
  \href{http://dx.doi.com/10.1080/10236190500138262}{DOI:
  10.1080/10236190500138262}.

\bibitem{Schneider:07d}
C.~Schneider, {\it {Simplifying Sums in $\Pi\Sigma$-Extensions}},  {\em J.
  Algebra Appl.} {\bf 6} (2007) 415--441,
  \href{http://dx.doi.com/10.1142/S0219498807002302}{DOI:
  10.1142/S0219498807002302}.

\bibitem{Schneider:08c}
C.~Schneider, {\it A refined difference field theory for symbolic summation},
  {\em J. Symbolic Comput.} {\bf 43} (2008) 611--644,
  [\href{http://xxx.lanl.gov/abs/arXiv:0808.2543}{{\tt arXiv:0808.2543}}].

\bibitem{Schneider:10a}
C.~Schneider, {\it {Structural Theorems for Symbolic Summation}},  {\em Appl.
  Algebra Engrg. Comm. Comput.} {\bf 21} (2010) 1--32,
  \href{http://dx.doi.com/10.1007/s00200--009--0115--3}{DOI:
  10.1007/s00200--009--0115--3}.

\bibitem{Schneider:10b}
C.~Schneider, {\it {A Symbolic Summation Approach to Find Optimal Nested Sum
  Representations}},  in {\em {Motives, Quantum Field Theory, and
  Pseudodifferential Operators}} (A.~Carey, D.~Ellwood, S.~Paycha, and
  S.~Rosenberg, eds.), vol.~12 of {\em Clay Mathematics Proceedings},
  pp.~285--308, Amer. Math. Soc, 2010.
\newblock \href{http://xxx.lanl.gov/abs/arXiv:0808.2543}{{\tt
  arXiv:0808.2543}}.

\bibitem{Schneider:10c}
C.~Schneider, {\it {Parameterized Telescoping Proves Algebraic Independence of
  Sums}},  {\em Ann. Comb.} {\bf 14} (2010) 533--552,
  [\href{http://xxx.lanl.gov/abs/arXiv:0808.2543}{{\tt arXiv:0808.2543}}].

\bibitem{Schneider:13b}
C.~Schneider, {\it {Fast Algorithms for Refined Parameterized Telescoping in
  Difference Fields}},  in {\em Computer Algebra and Polynomials} (M.~W.
  J.~Guitierrez, J.~Schicho, ed.), Lecture Notes in Computer Science (LNCS), to
  appear.
\newblock Springer, 2014.
\newblock \href{http://xxx.lanl.gov/abs/arXiv:1307.7887}{{\tt
  arXiv:1307.7887}}.

\bibitem{SIG1}
C.~Schneider, {\it Symbolic summation assists combinatorics},  {\em
  S\'em.~Lothar. Combin.} {\bf 56} (2007) 1--36, Article B56b,
  \href{http://www.emis.de/journals/SLC/wpapers/s56.html}{http://www.emis.de/journals/SLC/wpapers/s56.html}.

\bibitem{SIG2}
C.~Schneider, {\it Simplifying multiple sums in difference fields},  in {\em
  {Computer Algebra in Quantum Field Theory: Integration, Summation and Special
  Functions}} (C.~Schneider and J.~Bl\"umlein, eds.), Texts and Monographs in
  Symbolic Computation, pp.~325--360.
\newblock Springer, 2013.
\newblock \href{http://xxx.lanl.gov/abs/arXiv:1304.4134}{{\tt
  arXiv:1304.4134}}.

\bibitem{Ablinger:2010pb}
J.~Ablinger, J.~Blumlein, S.~Klein, and C.~Schneider, {\it {Modern Summation
  Methods and the Computation of 2- and 3-loop Feynman Diagrams}},  {\em
  Nucl.Phys.Proc.Suppl.} {\bf 205-206} (2010) 110--115,
  [\href{http://xxx.lanl.gov/abs/1006.4797}{{\tt 1006.4797}}].

\bibitem{Blumlein:2012hg}
J.~Blumlein, A.~Hasselhuhn, and C.~Schneider, {\it {Evaluation of Multi-Sums
  for Large Scale Problems}},  {\em PoS} {\bf RADCOR2011} (2011) 032,
  [\href{http://xxx.lanl.gov/abs/1202.4303}{{\tt 1202.4303}}].

\bibitem{Schneider:2013zna}
C.~Schneider, {\it {Modern Summation Methods for Loop Integrals in Quantum
  Field Theory: The Packages Sigma, EvaluateMultiSums and SumProduction}},
  {\em J.Phys.Conf.Ser.} {\bf 523} (2014) 012037,
  [\href{http://xxx.lanl.gov/abs/1310.0160}{{\tt 1310.0160}}].

\bibitem{Gluza:2007bd}
J.~Gluza, F.~Haas, K.~Kajda, and T.~Riemann, {\it {Automatizing the application
  of Mellin-Barnes representations for Feynman integrals}},  {\em PoS} {\bf
  ACAT2007} (2007) 081, [\href{http://xxx.lanl.gov/abs/0707.3567}{{\tt
  0707.3567}}].

\bibitem{Huber:2005yg}
T.~Huber and D.~Maitre, {\it {{HypExp}, a {Mathematica} package for expanding
  hypergeometric functions around integer-valued parameters}},  {\em Comput.
  Phys. Commun.} {\bf 175} (2006) 122--144,
  [\href{http://xxx.lanl.gov/abs/hep-ph/0507094}{{\tt hep-ph/0507094}}].

\bibitem{Huber:2007dx}
T.~Huber and D.~Maitre, {\it {HypExp 2, Expanding Hypergeometric Functions
  about Half-Integer Parameters}},  {\em Comput. Phys. Commun.} {\bf 178}
  (2008) 755--776, [\href{http://xxx.lanl.gov/abs/0708.2443}{{\tt 0708.2443}}].

\bibitem{Davydychev:2000na}
A.~I. Davydychev and M.~Y. Kalmykov, {\it {New results for the epsilon
  expansion of certain one, two and three loop Feynman diagrams}},  {\em
  Nucl.Phys.} {\bf B605} (2001) 266--318,
  [\href{http://xxx.lanl.gov/abs/hep-th/0012189}{{\tt hep-th/0012189}}].

\bibitem{Smirnov:2001cm}
V.~A. Smirnov, {\it {Analytical result for dimensionally regularized massive
  on-shell planar double box}},  {\em Phys.Lett.} {\bf B524} (2002) 129--136,
  [\href{http://xxx.lanl.gov/abs/hep-ph/0111160}{{\tt hep-ph/0111160}}].

\bibitem{Heinrich:2004iq}
G.~Heinrich and V.~Smirnov, {\it {Analytical evaluation of dimensionally
  regularized massive on-shell double boxes}},  {\em Phys. Lett.} {\bf B598}
  (2004) 55--66, [\href{http://xxx.lanl.gov/abs/hep-ph/0406053}{{\tt
  hep-ph/0406053}}].

\bibitem{Raab:LL2014}
J. Ablinger, J. Bl{\"u}mlein, C.G. Raab, C. Schneider, {\it Nested (inverse)
  binomial sums and new iterated integrals for massive Feynman diagrams}, talk
  held at Workshop {\it Loops and Legs in Quantum Field Theory}, April 27 - May
  2, 2014, Weimar, Germany; slides of talk at
  \href{https://indico.desy.de/conferenceDisplay.py?confId=8107}{LL2014
  webpage}.

\bibitem{Ablinger:2014bra}
J.~Ablinger, J.~Bl{\"u}mlein, C.~Raab, and C.~Schneider, {\it {Iterated
  Binomial Sums and their Associated Iterated Integrals}},
  \href{http://xxx.lanl.gov/abs/1407.1822}{{\tt 1407.1822}}.

\bibitem{Ablinger:2010kw}
J.~Ablinger, {\it {A Computer Algebra Toolbox for Harmonic Sums Related to
  Particle Physics}},  \href{http://xxx.lanl.gov/abs/1011.1176}{{\tt
  1011.1176}}.

\bibitem{Ablinger:2011te}
J.~Ablinger, J.~Blumlein, and C.~Schneider, {\it {Harmonic Sums and
  Polylogarithms Generated by Cyclotomic Polynomials}},  {\em J. Math. Phys.}
  {\bf 52} (2011) 102301, [\href{http://xxx.lanl.gov/abs/1105.6063}{{\tt
  1105.6063}}].

\bibitem{Ablinger:2013cf}
J.~Ablinger, J.~Bl{\"u}mlein, and C.~Schneider, {\it {Analytic and Algorithmic
  Aspects of Generalized Harmonic Sums and Polylogarithms}},  {\em J. Math.
  Phys.} {\bf 54} (2013) 082301, [\href{http://xxx.lanl.gov/abs/1302.0378}{{\tt
  1302.0378}}].

\bibitem{Ablinger:2013hcp}
J.~Ablinger, {\it {Computer Algebra Algorithms for Special Functions in
  Particle Physics}},  \href{http://xxx.lanl.gov/abs/1305.0687}{{\tt
  1305.0687}}.

\bibitem{Davydychev:2003mv}
A.~Davydychev and M.~Kalmykov, {\it {Massive {Feynman} diagrams and inverse
  binomial sums}},  {\em Nucl. Phys.} {\bf B699} (2004) 3--64,
  [\href{http://xxx.lanl.gov/abs/hep-th/0303162}{{\tt hep-th/0303162}}].

\bibitem{Remiddi:1999ew}
E.~Remiddi and J.~Vermaseren, {\it {Harmonic polylogarithms}},  {\em Int. J.
  Mod. Phys.} {\bf A15} (2000) 725--754,
  [\href{http://xxx.lanl.gov/abs/hep-ph/9905237}{{\tt hep-ph/9905237}}].

\bibitem{Ablinger:2013jta}
J.~Ablinger and J.~Bl{\"u}mlein, {\it {Harmonic Sums, Polylogarithms, Special
  Numbers, and their Generalizations}},
  \href{http://xxx.lanl.gov/abs/1304.7071}{{\tt 1304.7071}}.

\bibitem{panzer-hyperint}
E. Panzer, \verb|HyperInt| project webpage, Humboldt University, Berlin,
  \href{http://www.math.hu-berlin.de/~panzer/}{http://www.math.hu-berlin.de/~panzer/}.

\end{thebibliography}

\providecommand{\href}[2]{#2}\begingroup\raggedright\endgroup

\end{document}